\documentstyle[sprocl]{article}

\input{psfig}

\def\Journal#1#2#3#4{{#1} {\bf #2}, #3 (#4)}

\def\NPB{{\em Nucl. Phys.} B}

\def\be{\begin{equation}}
\def\ee{\end{equation}}
\def\bea{\begin{eqnarray}}
\def\eea{\end{eqnarray}}

\def\apj{\em ApJ}
\def\apjl{\em ApJ}

\def\aap{\em A\&A}

\def\iauc{\em IAU~Circ.}

\def\kpc {\, {\rm kpc}}
\def\kms {\,{\rm km \, s^{-1} }}
\def\kmsk {\,{\rm km \, s^{-1} \kpc^{-1}}}

\begin{document}

\title{Binary Microlening Event MACHO-98-SMC-1}  

\author{Sun Hong Rhie}

\address{Physics Department, University of Notre Dame, 
Notre Dame,\\ IN 46556, USA\\E-mail: srhie@nd.edu} 




\maketitle\abstracts{ The recent binary microlensing event toward
the Small Magellanic Cloud MACHO-98-SMC-1 was alerted by the MACHO
collaboration and monitored by many microlensing experiments for
its complete coverage of the second caustic crossing.   The purpose
of this global monitoring campaign was to determine the relative proper 
motion $|\vec\mu|$ of the lensing object with respect to the source star
that may be indicative of the location of the lensing object. (``Is it
a Galactic halo object or not, that is the question.") 
The estimated value $|\vec\mu| \approx 1.3 \kmsk$ indicates that the binary
lensing object belongs to the stellar population of the SMC.  
We discuss the implication of this binary event for the halo dark matter.
}

%

\section{Introduction}

Two and half years ago in the Santa Monica meeting, DM96, we~\cite{dm96}
 advocated that a microlensing planet search network can be used  for 
high time resolution observations of microlensing events toward the Magellanic 
Clouds {\it during the off-season of the Galactic Bulge} to help out dark 
matter search efforts.  MACHO-98-SMC-1 showed up as a microlensing event
(alerted by the MACHO collaboration May 25, 1998) 
{\it in the middle of the Bulge season}, 
and the Microlensing Planet Search collaboration (MPS)~\cite{98smc1-mps} 
immediately started monitoring the target star.  Microlensing events toward 
the Clouds are relatively rare (currently, $\sim 4$ events per year) and the 
necessity not to miss the possible chance to learn better about the lensing 
object is so much urgent.  On June 8, the event was announced to have crossed 
a caustic (first caustic crossing) by the MACHO/GMAN collaboration based 
on the data from CTIO 0.9m.  One of the MPS observations at June 5.55 UT 
would turn out to be crucial for the  reconstruction of the binary lens and
determining the relative proper motion $\mu$ of the lensing object.  

The estimation of the microlensing optical depth towards the Large 
Magellanic Cloud (LMC) by the MACHO collaboration~\cite{macho-lmc2},
$\tau  = 2.9^{+1.4}_{-0.9}\times 10^{-7}$, indicates that 
there are previously unknown ``dark lens populations" toward the LMC.  
If one assumes that the eight microlensing events toward the LMC are due
to lensing objects that constitute a fair sample of the halo mass density 
as described by the ``standard isothermal model" of the Galaxy, the 
distribution of the Einstein ring crossing time of the observed microlensing 
events, $20 < t_E < 70$ days ($<t_E> \sim 40$ days), indicates that the 
typical mass of the lensing objects is $\sim 0.5 {\rm M}_\odot$, which may be 
white dwarfs~\cite{wd}, or something more imaginative such as  primordial 
black holes (PBH) whose existence may be testable through the detection of 
gravitational waves generated by coalescing binary PBH's~\cite{bhole}  
or boson stars that can exist in scalar-gravity field theories.    

Or, the lensing objects may be normal stars in the Clouds as has been 
claimed by Sahu.  His estimation of the LMC self-lensing optical depth,
$\tau \sim 5\times 10^{-8}$, is ``far" short of the observed optical depth 
toward the LMC, however.  Gould~\cite{gould}  also showed that 
the LMC-LMC self-lensing optical depth  
is constrained by the line of sight velocity dispersion  and  estimated
the upper bound $\tau < 1\times 10^{-8}$.  
More recently, Weinberg~\cite{mweinberg} 
suggested a tidally flared LMC and a larger self-lensing optical depth.   
Since the hotter, the larger the velocity dispersion, it  will be interesting 
to see if the model is consistent with  the observed velocity dispersion of 
the LMC ($\sim 20 \kms$).   It will be more encouraging if the 2MASS 
studies reveal that the LMC disk is indeed extended as Weinberg predicts.  
(Figure 6 in Weinberg's shows that the 
``observed tidal radius" of the LMC $r_t \sim 10.8$ kpc is ``saturated" by 
$t = 0.5$ Gyr, and it is not clear if the rather stable evolution in the 
later epoch is an artifact of the tidal boundary condition.  Without the 
boundary condition, the numerical satellite galaxy may evaporate more 
freely.)   

There also have been suggestions that the ``dark lens population" 
be an intervening galaxy, tidal debris, a thick or warp component of the 
Galactic disk, etc.

It is the unsettling status of microlensing dark matter search experiments 
that the interpretation of their measurements is subject to  a variety 
of unknowns instead of setting global constraints on the intricate 
dynamics of matter.  
It is especially acute because in principle, one must be able to determine
the mass and the location of the lensing object for each microlensing event
within the microlensing measurements.  That is, if each microlensing event
is fully measured.  Lacking parallax satellites or astrometric measurements,
however, one can only determine a single quantity, namely, the Einstein ring 
radius crossing time $t_E$, out of three unknowns of distance, velocity 
and mass of the lensing object for most of the microlensing events 
-- the ``standard" or symmetric photometric microlensing events.     
Thus, the mass or the location of each individual microlensing object 
has never been directly measured, and the identity of the lensing objects
remains to be statistical which depends on the models of the Galaxy and
the Clouds.  Furthermore, the handful of microlensing events toward the 
Clouds (projected to be $\approx 20$ events by the end of 1999) are far too 
small a statistic to sort out the characteristics of the lensing populations
and determine whether there is the Galactic halo component.  The next  
generation  microlensing experiments~\cite{stubbs} may just come to 
the rescue with  hundreds of microlensing events toward the Clouds.  
Different lines of sight such as M31 and Fornax dwarf spheroid will have
to be actively investigated as well. 

In a caustic crossing binary lensing event, one can measure one more 
independent parameter, namely, the ``source radius crossing time", 
$t_{\ast}$,  and thereby estimate the relative proper motion $\mu$ 
of the lensing object with respect to the source star by independently 
determining the angular size of the  source star from its brightness and color.
The logic is that if $\mu$ is very small,  the binary lensing must be due 
to a self-lensing of the SMC.  So far, no one has come up with a halo model 
that allows $\mu$ as small as $\mu \sim 1 \kmsk$ with non-negligible 
probability.   For a typical halo lens we expect $\mu \approx 20 \kmsk$.
So, the semi-open global monitoring campaign was conducted involving 
most of the microlensing experiment teams.   $\mu \sim 1.3 \kmsk$ was  
small, and as a consequence, the monitoring continued more than 24 hours
in high time resolution and with high level of adrenaline.  One of the
conference attendee commented after the talk, ``It seems very interesting
even though chaotic."    The result was the first complete coverage of the 
second caustic crossing of the binary microlensing event MACHO-98-SMC-1.  
The details can be found in the reports issued by various collaborations:
the EROS collaboration~\cite{98smc1-eros}, 
the PLANET collaboration ~\cite{98smc1-planet},
the MACHO/GMAN collaboration~\cite{98smc1-macho},
the OGLE collaboration~\cite{98smc1-ogle},
and the MPS collaboration  ~\cite{98smc1-mps}.

\section{Observations and Reconstructions of the 
  Binary Microlensing Event MACHO-98-SMC-1}

The lensed star is located in the SMC at  
$(\alpha, \delta) = (00:45:35.2, \ 72:52:34.1)$ (J2000).   
The mass ratio of the binary lensing stars is $\approx 2:1$,
and their proejected distance is $\approx 2/3$ in units of 
the Einstein ring radius of the total mass.  (See table 1 of the MPS 
paper~\cite{98smc1-mps} for more details.)   
\begin{figure}[t]
\psfig{figure=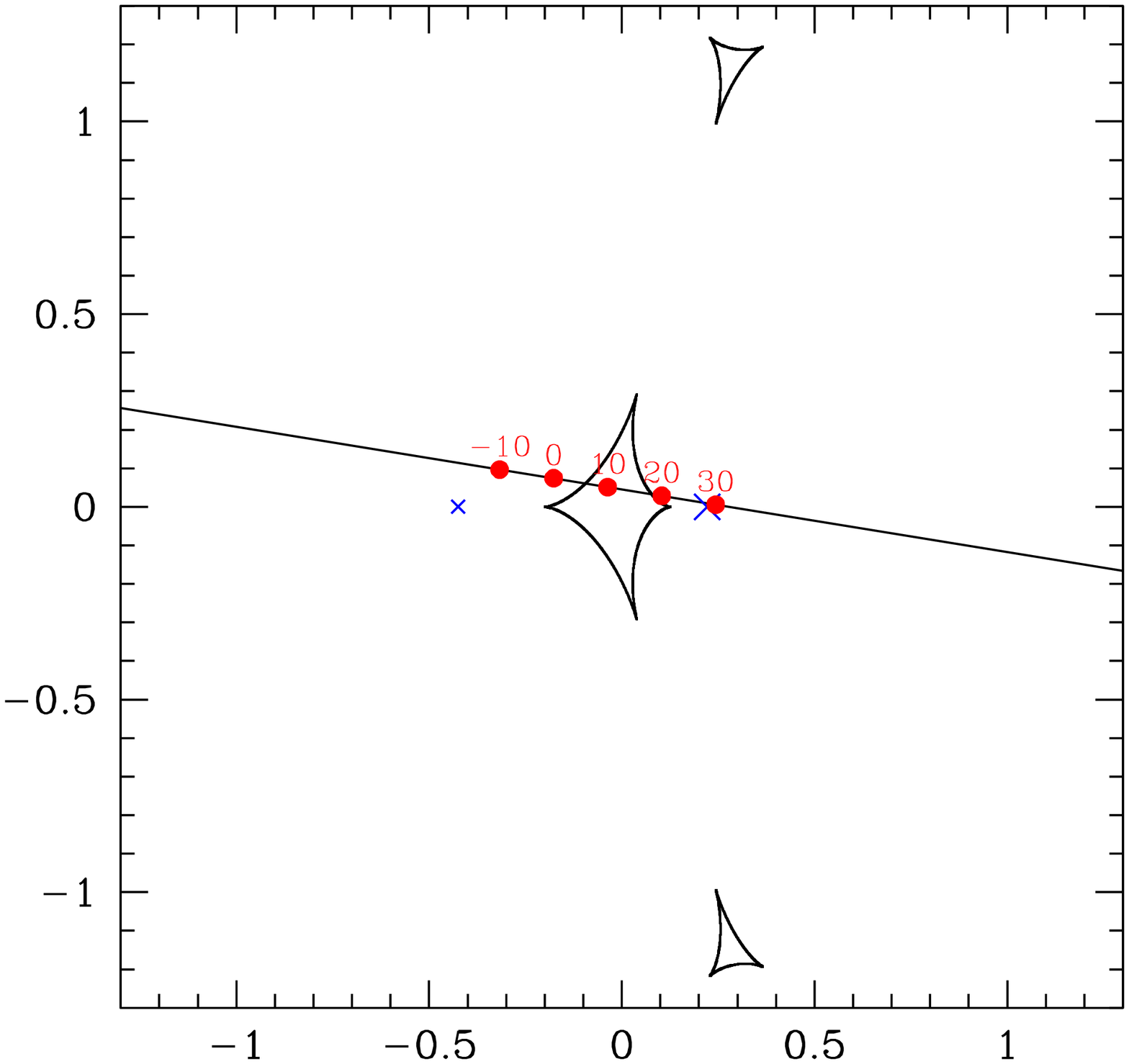,height=5in}
\caption{This figure shows the configuration of the caustic curves for the
MPS lightcurve fit to binary lensing event MACHO-98-SMC-1. The crosses
indicate the locations of the lenses, and the straight line indicates the
path of the source star with respect to the caustic curves. The red dots on
the source star path indicate the location of the source at various dates
given in June, UT. The distance scale for the axes is the Einstein ring
radius, $R_E$. Note that the actual size of the source star is only about
$0.0015R_E$ that is much less than the thickness of the curves in the
Figure.  \label{fig:caustic}}
\end{figure}
Figure ~\ref{fig:caustic} shows the trajectory of the source across the
caustic curve where the travese took about 12 days.  

The world-wide monitoring campaign was mostly focused on the coverage of 
the second caustic crossing where the lensed star exited the caustic curve.
The peak turn-over occurred around June 18.0 UT over South Africa, and
it was observed from SAAO 1m by the PLANET collaboration. Also, the 
spectrum was taken at the peak, and the star was determined to be a
A star with $T \approx 8000$K. 
\begin{figure}[t]
\psfig{figure=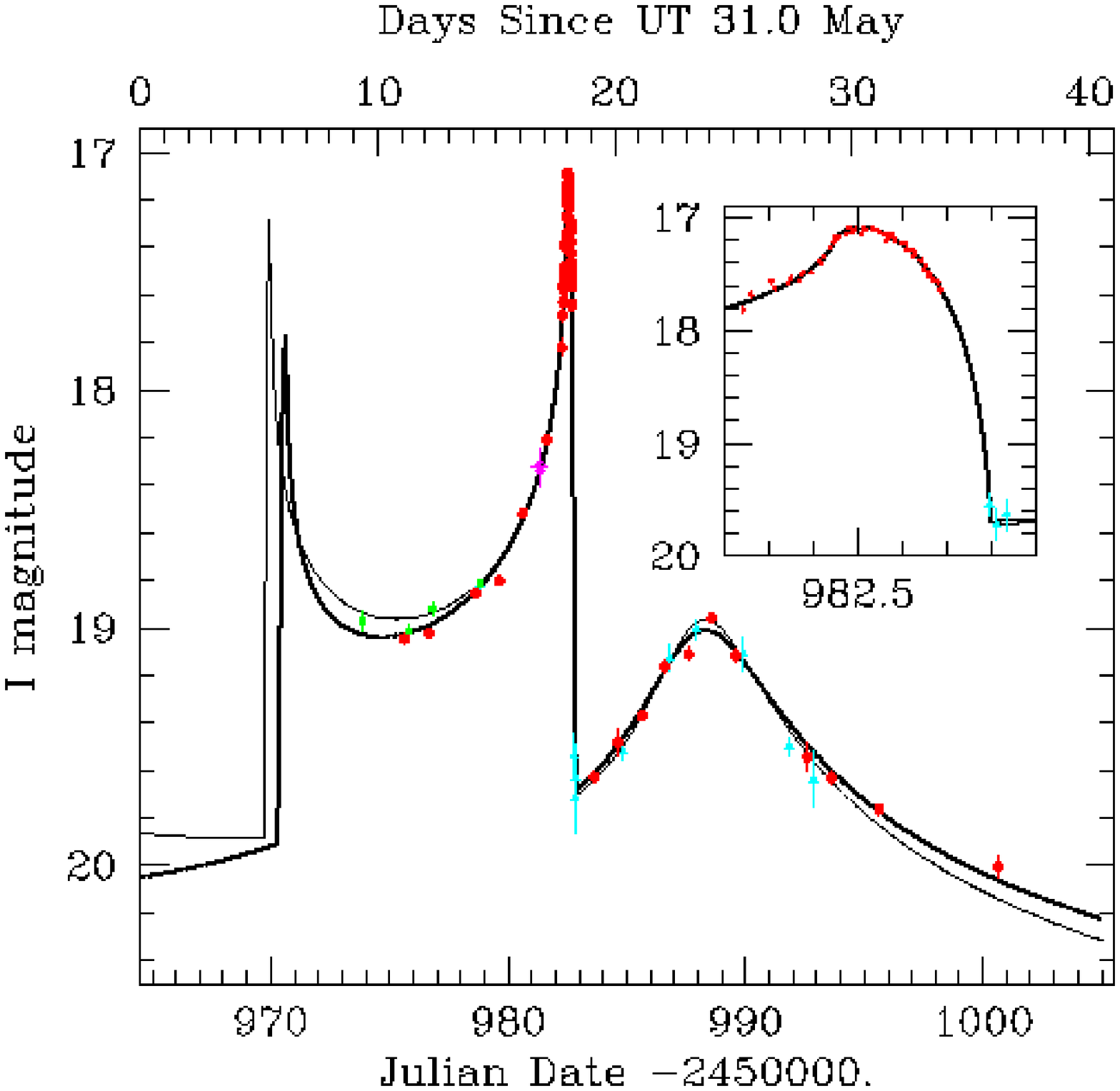,height=4.9in}
\caption{This figure shows the photometric data and fit curves of 
the binary microlensing event MACHO-98-SMC-1 from the article by 
the PLANET collaboraion.   The inset shows the
peak turn-over data from South Africa.  The curvature change of the light 
curve around $t \approx 982.43$ is where the limb of the lensed star 
``touches" the caustic curve from inside.  Here the apparent curvature 
change looks more prominent because of guding fitting curve for which
the homogeneous luminosity profile of the lensed star was used.   The two 
fit curves presented by the PLANET collaboration illustrate the kind of 
uncertainties in the reconstruction of binary lensing light curves from 
``incomplete data."  The one with the earlier first caustic crossing will 
be ruled out by the MPS observation at June 5.55 UT.   
\label{fig:peak}}
\end{figure}
See figure ~\ref{fig:peak} for the PLANET data and their binary lens
fit light curves.  The high resolution peak turn-over data from South
Africa can be seen in the inset.
The curvature change of the light curve around $t \approx 982.43$
is where the limb of the lensed star "touches" the caustic curve from 
inside.   Here, the apparent curvature
change looks more prominent because of the homogeneous luminosity 
profile (constant surface luminosity over the disk of the star) used 
for the fit curve.   
The two fit curves disagree notably near the first caustic crossing 
and ealier where the PLANET data coverage is null.  The binary lens model
with earlier first caustic crossing  resulted in $\mu \sim 2 \kmsk$ 
offering  the possibility  that the lensing binary belongs to a foreground 
tidal debris~\cite{98smc1-planet} or  the Galactic halo~\cite{honma}.    
MACHO/GMAN~\cite{98smc1-macho} found this fit to be inconsisent with 
their baseline data prior to the caustic crossings.
Now it is ruled out by the MPS observation at June 5.55 UT 
that is consistent with a slowly rising amplification of a source star
that approaches a binary caustic.   The first caustic crossing occurred
around June 6.0 UT over the South Africa.   

In a binary lensing, the number of images are three or five: three
``normal" images when the source is outside the caustic loops and
two extra images inside.  The ``normal" images are always full images.
The two extra images are partial and are connected across the critical
curve, however, when the source star crosses the caustic curve. 
The critical curve is where the magnification of a point source diverges.  
Thus, the two extra images are very bright.  As the source exits the caustic,
the two partial conjoined images completely disappear (``into the critical
curve"), and the star dims rapidly.   This rapid change was observed by 
the EROS~\cite{98smc1-eros}.   
This was the first time the linear characteristic
of the falling light curve of a caustic crossing was actually measured.
\begin{figure}[t]
\psfig{figure=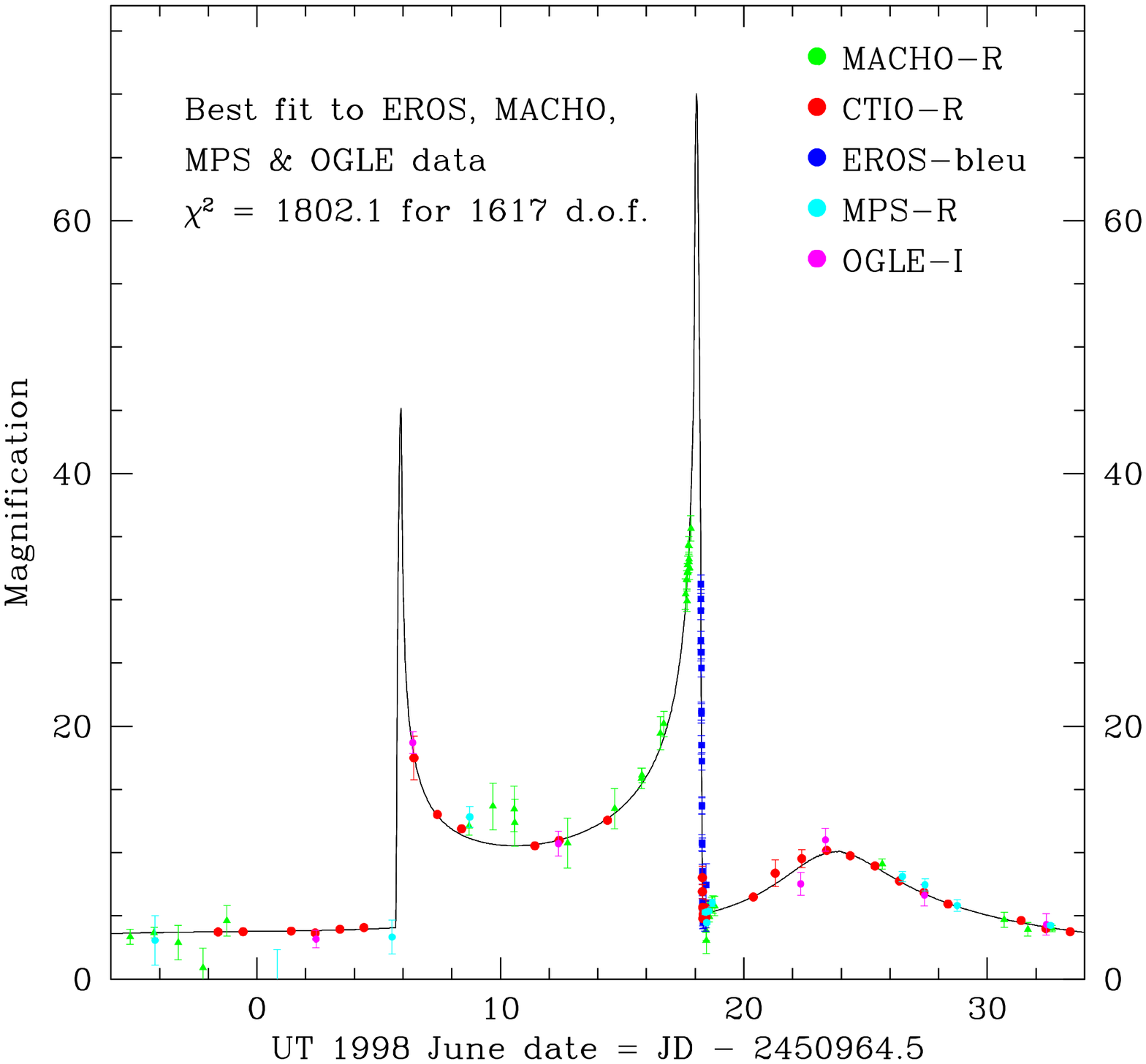,height=5in}
\caption{This figure shows the MPS lightcurve fit to binary lensing event
MACHO-98-SMC-1 as represented by the data from the EROS, MACHO/GMAN,
MPS and OGLE collaborations.   
\label{fig:lcurve}}
\end{figure}

Figure ~\ref{fig:lcurve} shows the photometric data from EROS, MACHO/GMAN, 
MPS and OGLE collaborations and the MPS lightcurve fit to the data. 
While the PLANET data were unavailable,  it was an opportune case to test
if we could find the correct binary lens parameters and reconstruct
the ``missing peak" from the joint data shown in figure ~\ref{fig:lcurve}.   
The MPS analysis team found that the fitting process was arduously slow
without the MPS data constraining the first caustic crossing in time.    
According to the MPS fit,  the maximum amplification of the second 
caustic crossing was $\sim 70$. The PLANET fit with later first caustic
crossing produced the maximum amplificaiton $\sim 100$, for which the
blending level was ``preassigned" perhaps due to the lack of baseline
data.   Thus, the work of the PLANET 
collaboration amounted to a ``complementary test" whether the correct
binary lens can be reconstructed from the high resolution data for the
second caustic crossing (plus the MACHO/GMAN data that were made
available with the MPS real time fitting during the monitoring campaign: 
http://darkstar.astro.washington.edu/  and 
http://bustard.phys.nd.edu/MPS/.)

Caustic crossing binary lensing event is an exciting object all by itself.
To the eyes of beholders, the spectacular increase of the light flux
during the caustic crossing can be as cathartic as witnessing the resonance 
of  the charm-anti-charm bound state.  However, the real motivation 
had derived from the relatively rare opportunity to measure the 
relative proper motion of the lensing object~\cite{iauc6935}~\cite{iauc6939}
and throw 
a clean verdict whether the lensing object is in the Galactic halo or not.   
A Galactic halo object will have a high proper motion, while the proper
motion of an SMC object will be small.  The ``long duration" brilliance of 
the caustic crossing that was watched from  all three continents in the Southern 
hemisphere  turned out to mean a slow progression of the star across 
the caustic and the conclusion that the lensing object is not in the 
Galactic halo.  The size of the star was determined to be $\approx 1.1$
solar radius, and $\mu \approx 1.3 \kmsk$.    

There have been claims~\cite{dominik} that binary microlensing events 
can not be reconstructed uniquely.   These claims were made without any
analysis of the correlation with the quality of the data as if it were a 
generic feature of binary lensing.   The case MACHO-98-SMC-1 restores the
common sense: in a plagiarism of James Carville, ``Coverage, stupid!"  
Obviously, if we have one datum on the light curve,  there will be 
infinitely many other light curves that fit the datum.  There is a need
for systematic analysis for ``smart coverages".     

In fact, the coverage of the MACHO-98-SMC-1 is judged to be good enough 
to test the limb darkening profile of the lensed star.  This is planned
for the ``grand joint fit" where all the data will be combined and 
analysed. 
\begin{figure}[t]
\psfig{figure=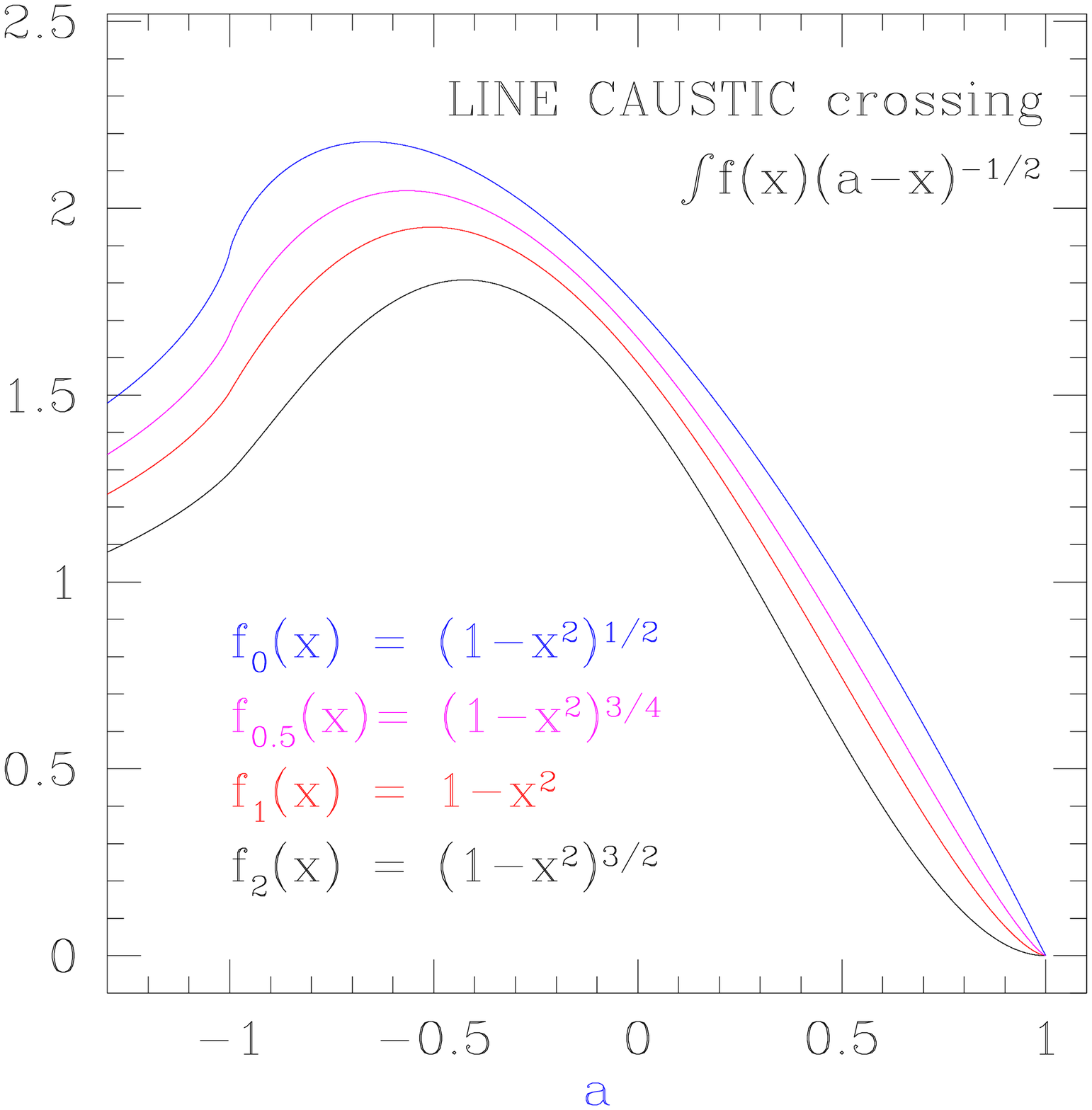,height=5in}
\caption{As the caustic line scans across the surface of the star, 
the integrated surface luminosity profile is ``read out" in time. 
This figure shows the behavior of the analytic terms: 
homogeneous ($f_0$), square root ($f_{0.5}$)., linear ($f_1$) 
and quadratic ($f_2$).  The star ``touches" the caustic line at $a = -1$
and exits the caustic at $a = 1$ in this plot. 
\label{fig:limb}}
\end{figure}
Figure ~\ref{fig:limb} shows the (unnormalized) second caustic crossing
light curves of the extra two images of stars with different luminosity 
profiles~\cite{limb}.  

\section{What can we conclude?}

MACHO-98-SMC-1 is the second microlensing event discovered toward the
SMC, is a binary lensing event, and is an SMC-SMC event.  
What does it tell us?  It is a peculiar reminder that the first LMC
event discovered by the MACHO collaboration was most likely a binary
lensing event~\cite{dm96}, and the   LMC microlensing statistic is dominated 
by single lens events.  Binary objects are known for stellar populations. 
Black hole Machos are also expected to in binaries~\cite{bhole} at a 
less level of $\approx 8\%$  even though an accurate estimation will need 
more sophisticated  calculations.  Boson star binary statistics are not 
available in literature.   SMC has been expected to have a high 
self-lensing probability because of the elongated feature toward us.   
So, what can we conclude?  Much more statistics and active monitoring
of microlensing events for extra information besides the Einstein ring
crossing time!!!



\newpage

\begin{thebibliography}{99}

\bibitem{dm96}S. Rhie and D. Bennett,
            \Journal{\NPB(Proc.Suppl.)}{51B}{86}{1996}.

\bibitem{98smc1-mps}S. Rhie {\it et al}
            (Microlensing Planet Search collaboration) 1998,
            \Journal{\apj}{}{Submitted}{1998},
             astro-ph/9812252.

\bibitem{macho-lmc2}C. Alcock {\it et al},
           \Journal{\apj}{486}{697}{1997}.

\bibitem{wd}G. Chabrier, \Journal{\apjl}{}{Accepted}{1999},
             astro-ph/9901145.

\bibitem{bhole}T. Nakamura {\it et al},
            \Journal{\apjl}{487}{L139}{1997}.

\bibitem{sahu}K. C. Sahu, \Journal{\em Nature}{370}{275}{1995}.

\bibitem{gould}A. Gould, \Journal{\apj}{441}{77}{1995}.

\bibitem{mweinberg}M. Weinberg, 1998, astro-ph/9811204

\bibitem{stubbs}C. Stubbs, 1998, astro-ph/9810488

\bibitem{98smc1-eros}C. Afonso {\it et al}, 
          \Journal{\aap}{337}{L17}{1998}.

\bibitem{98smc1-planet}M. Albrow {\it et al}, 
           \Journal{\apjl}{}{}{1999},  
            astro-ph/9807086.

\bibitem{98smc1-macho}C. Alcock {\it et al}, 
           \Journal{\apj}{}{Submitted}{1998}, 
            astro-ph/9807163. 

\bibitem{98smc1-ogle}A. Udalski {\it et al},
            \Journal{\em Acta Astronomica}{}{Submitted}{1998},
             astro-ph/9808077.

\bibitem{honma}M. Honma, 
               astro-ph/9811397

\bibitem{iauc6935}A. Becker {\it et al},
            \Journal{\iauc}{6935}{1}{1998}. 

\bibitem{iauc6939}D. Bennett {\it et al}, 
            \Journal{\iauc}{6939}{1}{1998}. 

\bibitem{dominik}M. Dominik, \Journal{\em A \& A}{}{Accepted}{1998}, 
             astro-ph/9703003

\bibitem{limb}S. Rhie, 
   {\it ``Limb Darkening and Line Caustic Crossing Microlensing"} (1998),
               astro-ph/9999999


\end{thebibliography}
\end{document}